# An alternative strategy for the use of a low-cost age-hardenable Fe-Si-Ti steel for automotive application

O. Bouaziz[1,2]
[1]*Laboratoire d'Etude des Microstructures et de Mécanique des Matériaux (LEM3), CNRS, Université de Lorraine, Arts et Métier Paris Tech, F 57000 , Metz , France*
[2]*LABoratoire d'EXcellence DAMAS*

corresponding author : O. Bouaziz
email : olivier.bouaziz@univ-lorraine.fr
phone number : 0033 626344383

## Abstract

For High Strength Low Alloy (HSLA) steels or for age-hardenable steels (maraging) the strengthening by precipitation is done before forming operation in order to increase the yield stress as much as possible. In this publication the advantages of an hardening thermal treatment after forming operation are investigated in a low cost age-hardenable steel Fe-Si-Ti consistent with automotive application.

Keywords : age-hardenable, precipitation, forming, ductility

## Introduction

The high strength low alloy (HSLA) steels are a group of low carbon steels with small amounts of alloying elements (as Ti, V, Nb, etc.) to obtain a good combination of strength, toughness and weldability [1–3]. By the addition of micro-alloying elements (less than 0.5 wt%), HSLA steels are strengthened by grain refinement strengthening, solid solution strengthening and precipitation hardening [4–8]. Regarding automotive applications, the entire range of HSLA steels are suitable for structural components of cars, trucks and trailers such as suspension systems, chassis, and reinforcement parts. HSLA steels offer good fatigue and impact strengths. Given these characteristics, this class of steels offers weight reduction for reinforcement and structural components [9-11].

Despite the interest of HSLA, the precipitation hardening is about 100MPa. So the new targets concerning the $CO_2$ emission of vehicles push the steel-makers to develop Advanced High Strength Steels (Dual-Phase, Transformation Induced Plasticity) hardened by multiphase microstructures containing 10 to 100% of martensite [12] which offer a better combination between strength and ductility for an acceptable cost Let's notice that the Ultimate Tensile Strength (UTS) increases more that the Yield Stress (YS). YS is crucial for anti-intrusive aspect during a crash [12].

In the other side, because there is no phase transformation in Aluminium, age-hardening by precipitation is widely used in Aluminium alloys hardened by various additive elements [see 13 for a review]. The volume fraction of precipitates is several percents, the hardening can be increased up to 400MPa but the ductility decreases rapidly.

In steel, there is only maraging steels which are strengthened by a massive precipitation of a martensitic matrix [see 14 for a review]. Despite the impressive YS (up to 2.5GPa), the uniform elongation is less that 1%. Consequently no forming operation is possible. In addition the high contents of nickel (about 18%), of molybdenum and of cobalt make these steels very expensive and so they are never used in automotive industry. On the contrary, in the 60's authors investigated other kind of steels suitable to be strongly hardened by massive intermetallic precipitation without extra-cost [15,16]. Among the different investigated system the Fe-Si-Ti alloys are the most promising. This is the reason why a serie of publications has been dedicated to this system in the 70's by Jack&al. [17-19]. Unfortunately hardness and compression behaviour have been only reported and a lot of discussions concerned the nature of the precipitates ($Fe_2Ti$, $FeSi$, or $Fe_2SiTi$) have been reported

More recently a systematic study of precipitation kinetics in a Fe-2.5%Si-1%Ti alloy in the temperature range 723 K to 853 K (450°C to 580°C), combining complementary tools (transmission electron microscopy (TEM), atom probe tomography (APT), and Small-Angle-Neutron Scattering (SANS)) have been carried out [20]. It has been shown that the Heusler phase $Fe_2SiTi$ dominates the precipitation process in the investigated time and temperature range, regardless of the details of the initial temperature history.

Taking into account that the ductility decreases regularly as a function of the hardening up to 1.2GPa and because it is targeted to obtain a steel suitable for deep drawing the strategy showed in Fig1. (orange arrow) has been investigated. The objective is to form a material as soft as possible and as ductile as possible and to obtain the hardening by a thermal treatment after forming. This publication presents the characterization of this way to manage the situation. It is noticed that as Si and Ti promote ferrite, there is no phase transformation whatever the thermal treatment.

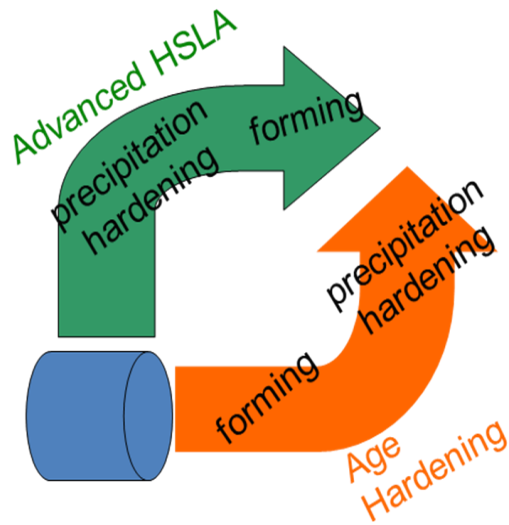

Fig1. Summary of the different strategies for the use of precipitation hardened steels : The green arrow is the usual one, the orange is the strategy developed in this publication

## Results

In order to assess for the first time in steel for automotive applications, the hardening have been chosen followed a thermal treatment at 500°C during two hours consistent with the kinetic determined by SANS [20]. As shown in Fig2. illustrating the tensile behaviour of the steel consisting only in solid solution (i.e. only after recrystallization) is very ductile as for IF steels but with a higher yield stress due to solid solution hardening. After treatment an hardening of 300MPa is obtained with promising ductility. As shown in Fig.3 the treatment have induced the expected massive precipitation of $Fe_2SiTi$ (3.8% weight percent with a radius of 4nm determined by TEM and SANS [20]).

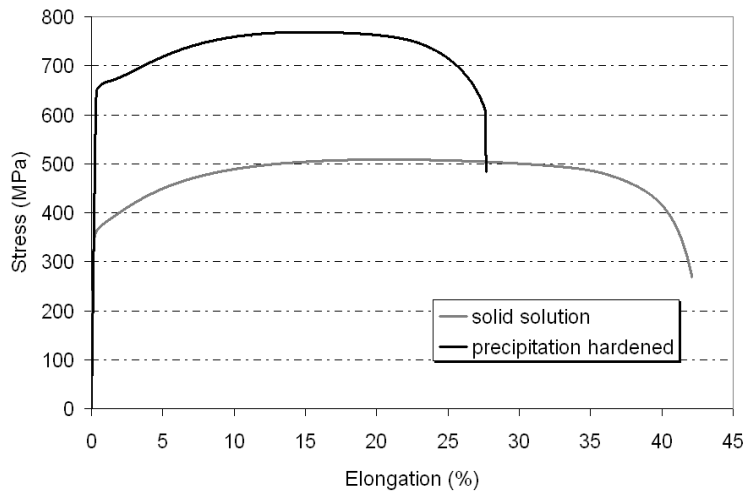

Fig2. Tensile curves before and after the ageing treatment

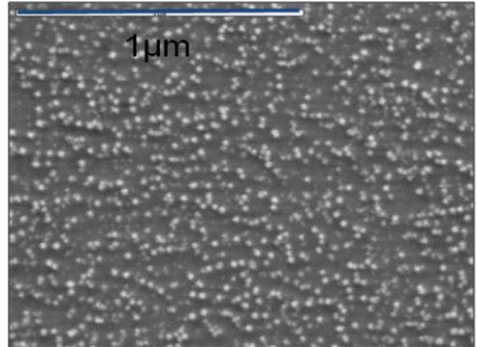

Fig 3. : TEM micrography showing the massive precipitation of $Fe_2SiTi$ after 2 hours at 500°C (the composition have been determined by APT [20])

In order to assess the metallurgical route, the Fig4. shows that it is possible to severely bend the steel consisting only in solid solution without any defect up to an angle of 180°. Hardness trough the thickness has been measured before and after the thermal treatment (Fig5.). The value confirm the hardening by precipitation. It is noticed that tis precipitation hardening is not sensitive to the strain hardening induced by bending trough the thickness.

Because the alloy is dedicated to automotive industry, the drawing has to be investigated. This the reason why a cup obtained by deep drawing of 5cm diameter has been manufactured using the steel consisting in solid solution without any problem as illustrated in Fig.6. It confirms the very high ductility before heat treatment.

One another problem in automotive industry is the increase in spring-back with an increase in strength. In addition it is very difficult to predict or to model this phenomenon. As shown in Fig7. this aspect has been studied by a standard test base the forming of a hat-shaped part. It is highlighted that the spring-back during the treatment is weak. That is probably because there is no phase transformation during precipitation and so no internal stresses.

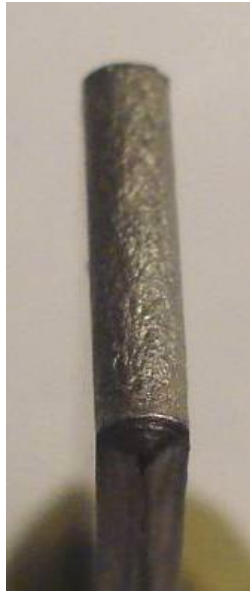

Fig4. Fully bent specimen before heat treatment (bending angle of 180°)

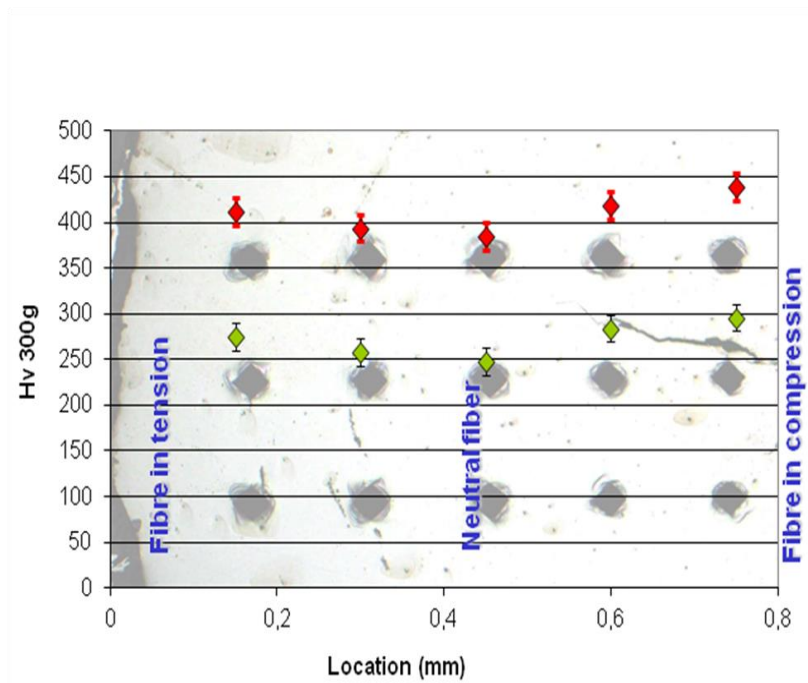

Fig5. Hardness measurement trough the thickness of the fully bent specimen (i.e. angle of 180°) before and after heat treatment

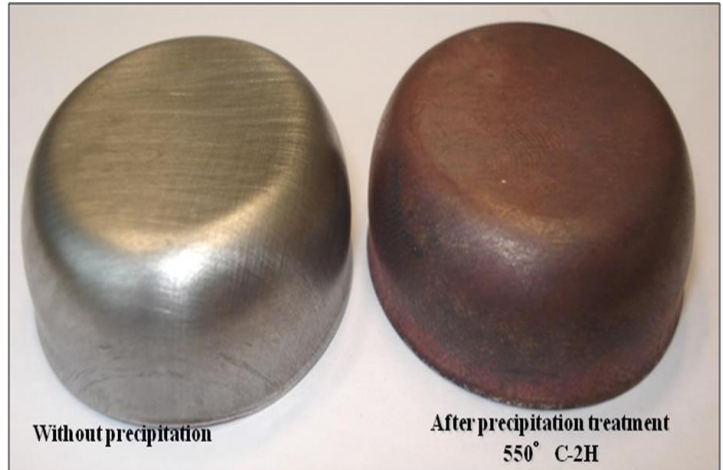
Fig6. Cup drawing before the hardening heat treatment (5cm diameter)

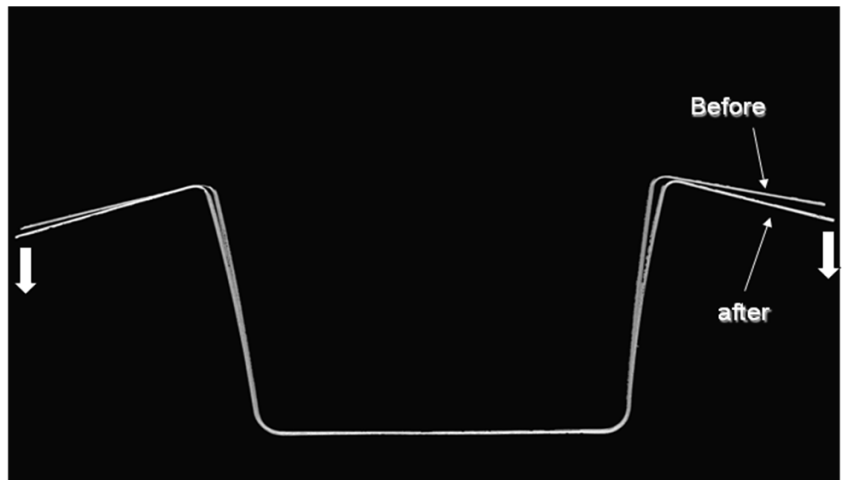
Fig 7. Evaluation of the effect of heat treatment on the spring-back

## Conclusion

For the first time in steel industry for automotive industry a low cost age-hardenable steel have been studied following a strategy based on forming operations before the heat treatment. The bending, the drawability and the spring-back have been investigated highlighting promising results. In addition the alloy exhibits a cost acceptable for automotive industry. In the future crash-worthiness and weldability should be assessed after heat treatment. One of the last advantage is that a lot of parts can be treated in the same time in a furnace usually dedicated to tempering treatment.